# Dual Stage Classification of Hand Gestures using Surface Electromyogram

Karush Suri*, Rinki Gupta

Electronics and Communication Engineering Department
Amity University Noida, Uttar Pradesh-201313, India
karushsuri@gmail.com, rgupta3@amity.edu

*Abstract*—Surface electromyography (sEMG) is becoming exceeding useful in applications involving analysis of human motion such as in human-machine interface, assistive technology, healthcare and prosthetic development. The proposed work presents a novel dual stage classification approach for classification of grasping gestures from sEMG signals. A statistical assessment of these activities is presented to determine the similar characteristics between the considered activities. Similar activities are grouped together. In the first stage of classification, an activity is identified as belonging to a group, which is then further classified as one of the activities within the group in the second stage of classification. The performance of the proposed approach is compared to the conventional single stage classification approach in terms of classification accuracies. The classification accuracies obtained using the proposed dual stage classification are significantly higher as compared to that for single stage classification.

*Keywords—Surface EMG; Multistage classification; PCA; SVM classifier; MANOVA; Dendogram*

## I. Introduction

Designing an efficient system capable of distinguishing the various hand gestures is of significant importance in applications ranging from consumer electronics such as human machine interfaces to sport sciences and elderly health care. A vast scope is witnessed when working with amputees where a comprehensive use of motion analysis is required [1]. Muscle motor functionality generates electrical signals in human body, which may be recorded using sensors placed in the surface of skin. These surface electromyogram (sEMG) signals can then be further processed and analyzed in order to solve diagnostic problems. The non-invasive nature of the technology makes it particularly useful in consumer electronics.

Directly analyzing raw sEMG signals for classification of hand gestures would require considerably large memory and computation time. Hence, the sEMG signals are first represented in terms of suitable feature set. Several time-domain, frequency domain [2], joint time-frequency domain [3] and higher order statistical features [4] are available in literature for analysis of sEMG signals. The feature set should be such that the underlying gestures are distinguishable from one another, while the features are similar for a number of repetitions of the same gesture. The distinction of different hand gestures requires a number of features which can be reduced in order to yield the most optimum values. Principle Component Analysis (PCA) and the Uncorrelated Linear Discriminant Analysis (ULDA) are the two suitable feature reduction schemes [3]. These two schemes can be contrasted between depending upon the number of features reduced or the quality of the features selected [2]. An important advantage of PCA over ULDA is that even there is significant correlation between the reduced feature set and the hand gestures and the confusion effect is reduced as a result of the orthogonal geometry of the simplified datasets.

Once the number of features is simplified, the classification of the hand gestures may be carried out using the classifiers available in literature. These include techniques such as Linear Discriminant Analysis, k-Nearest Neighbors, Multi-Layer Perceptrons and Support Vector Machines [1]. Support Vector Machines (SVM) with a non-linear kernel function has been reported to provide better classification of hand activities in terms of classification accuracies, as compared to other classifiers [1]. However, the approach prominently described in literature involves the classification of several hand activities in a single stage. This conventional approach leads to a higher complexity in classification since features that may be optimum for distinguishing between a few activities may not be suitable for another set of activities, and hence, such features must be discarded from the feature set during classification of all activities.

A more logical approach would be to divide the classification problem into several stages, wherein certain activities may be grouped together under one category. First the classification may be performed to identify the group to which the activity being classified may belong to using one set of features. Subsequently, the activity may be identified from the group of activities using another set of features. Such multi-stage classification approach has been demonstrated with speech signals [5] and also sEMG based hand gesture classification [6]. In [5], the authors have presented a clear contrast in the emotions of the speaker by separating the dissimilar entities first. In [6], hand activities such as finger movement and two grasping activities are first classified as separate groups from which the individual activities are then classified.

In this paper, a dual stage classification approach is proposed for classification of six grasping gestures of hand using sEMG signals recorded from two channels on the forearm. The suitability of the activity groups considered in the first stage of the proposed algorithm is demonstrated using a statistical measure. The performance of proposed dual-stage classification of grasping gestures is compared with the conventional single stage classification approach in terms of classification accuracies. The remaining paper is organized as follows. Section II contains the description of the features considered for classification and a brief description of the feature reduction and classification techniques used in this work. The proposed dual stage classification algorithm is presented in Sec. III. Results depicting the suitability of proposed grouping of the grasping activities and the classification of six grasping activities using actual sEMG signals are given in Sec. IV. Section V concludes the paper.

## II. CONVENTIONAL APPROACH FOR HAND ACTIVITY CLASSIFICATION

### A. The sEMG dataset

The sEMG signals considered in this work have been recorded from the surface of skin over the muscles Flexor Capri Ulnaris and Extensor Capri Radialis brevis [3], as depicted in Fig. 1. The objective is to utilize the sEMG signals to classify six grasping activities, listed in Table I. Table I also contains the class indicators for each activity, which are used for representing the activity in the analyses that follows as well as a pictorial description of the activities. The sEMG data has been collected from 5 healthy subjects, 3 female and 2 male, all subjects in the age group of 20-22 years. Grasps have been performed for 6 seconds with 30 repetitions for each activity. Use of limited number of sEMG sensors not only helps in reducing the cost of the sEMG based wearable device, but also makes it more wearable. The features considered in this paper for analysis of sEMG signals are summarized as follows.

### B. Features for analysis of sEMG signals

The most prominently used time-domain features are considered here [2]. These include:

*1) Mean absolute value (MAV)*: Mean absolute value (MAV) is the absolute value of the mean of a segment of data. MAV of a random variable $x$ having $n$ values can be expressed mathematically as

$$\text{MAV} = \frac{1}{n}\sum_{i=1}^{n} x_i. \quad (1)$$

*2) Standard Deviation*: Standards deviation is defined as the square root of the variance

$$\sigma = \sqrt{\frac{1}{n-1}\sum_{i=0}^{n}(x_i - \mu)^2}, \quad (2)$$

where $\mu$ is the mean of the data.

*3) Root Mean Square (RMS)*: The RMS value of a data segment is the root of the mean of squares of the data values. The RMS of a random variable $x$ having $n$ values can be expressed mathematically as

$$\text{RMS} = \sqrt{\frac{1}{n}\sum_{i=0}^{n} x_i^2}. \quad (3)$$

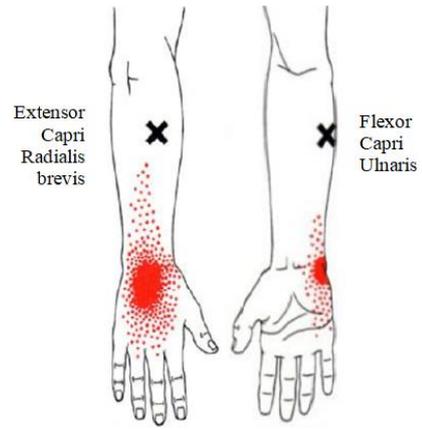

Fig. 1. Muscles monitored for acquiring sEMG signals

TABLE I. CLASS DETAILS FOR THE SIX ACTIVITIES

| Name of the activity | Class Indicators | Description |
|---|---|---|
| Palmar | P | |
| Lateral | L | |
| Tip | T | |
| Hook | H | |
| Spherical | S | |
| Cylindrical | C | |

*4) Slope Sign Change*: It counts the number of instances where the slope of the signal changed its sign. For any three EMG amplitude samples, $x_{k-1}, x_k$ and $x_{k+1}$, where $k$ is any positive integer, the number of slope sign changes is given by $\sum f(x)$, where $f(x)$ is

$$f(x) = \begin{cases} 1, & \text{if } x_k < x_{k+1} \text{ and } x_k < x_{k-1} \\ 0, & \text{if } x_k > x_{k+1} \text{ and } x_k > x_{k-1} \end{cases}. \quad (4)$$

*5) Waveform Length(WL)*: It is a cumulative variation of the sEMG that depicts the degree of variation about the signal, given as

$$\text{WL} = \sum_{k=1}^{n-1} |x_{k+1} - x_k|. \quad (5)$$

*6) Auto-Regressive Coefficients:* These are the constant coefficients $(a_r)$ of the auto-regressive (AR) model $\hat{x}$ of the sampled instants of the sEMG signal, given as

$$\hat{x}_n = \sum_{k=1}^{p} a_k x_{n-k} + \varepsilon, \quad (6)$$

where $p$ is the order of the AR model and $\varepsilon$ is the prediction error.

*7) Skewness:* The skewness of any signal segment with variable value $x$, having mean μ and standard deviation σ is given as

$$\gamma = \frac{E\{x-\mu\}^3}{\sigma^3}, \qquad (7)$$

where $E$ is the expectation operator.

*8) Kurtosis:* The kurtosis of any signal segment with variable value $x$, having mean μ and standard deviation σ is given as

$$\kappa = \frac{E\{x-\mu\}^4}{\sigma^4}. \qquad (8)$$

*C. Feature reduction and classification*

Dimensionality reduction leads to an improvement in the classification performance due to reduction in correlation in the feature set. Using the features mentioned above in the PCA based feature selection process, the number of features is experimentally reduced to a minimum while maintaining the classification accuracies as high as possible [7]. The PCA algorithm projects the *d*-dimensionality data onto *l* eigenvectors corresponding to their covariance matrix, thus leading to a linear transformation of the original space. The PCA algorithm computes the principle arguments in two mutually orthogonal directions by simplifying the multivariate data. Reduction in the dimensionality serves as an important step in order to predict the values for multiple observations with similar characteristic traits.

The SVM maps the inputs implicitly to its corresponding outputs by making use of the most optimized hyper-plane in the high- dimensional feature space [8]. Non-linear classification in SVM makes use of the kernel trick where a kernel, which is a generalized form of the function is used. This enables the function to map the entities in a high dimensional feature space by successive computation of the dot product. In case of binary classification, the predicted values are consequence of the two classes. However, in case of six independent activities the method involves the use of masking the residues. Each activity is considered as a separate entity and then distinguished from the remaining counterparts. The replicates contained in the test data are predicted on the basis of which the accuracy of this single stage classification is assessed.

On analyzing the results for single stage conventional method (discussed in Section IV), further scope of improvement lies in first classifying the dissimilar activities as a group. This would help in avoiding confusion given a set a features for classification. Here, grouping is proposed to be carried out on the basis of common traits, such as power and precision grasping [9]. Certain activities require a greater use of the muscle power while, some activities need a precise idea regarding the location of the object. These groups can then be classified further in the later stages, as explained in the following section.

## III. PROPOSED DUAL STAGE HAND ACTIVITY CLASSIFICATION

In this work, classification of six grasping activities are considered, which are given in Table I. As depicted in Fig. 2, in the conventional approach available in literature, the six activities are considered as individual classes and the classifiers attempts to classify given sEMG signal as originating from one of the six activities following the feature extraction and the feature reduction steps. Alternately, the grasping activities may be broadly categorized into two groups according to the degree of freedom (DoF) required to perform the activities. Certain tasks require significant amount of power such as lifting and stretching, while some other tasks are based on the use of precision and accuracy such as holding and writing [9]. A detailed vector analysis along with inverse kinematics of the DoFs has been carried out in [9].

Consequently, in the proposed dual-stage classification, the considered activities are grouped on the basis of two characteristics, power and precision. Among the considered activities, similar tasks involving the use of power are grouped together in the corresponding group in the first stage. These consist of cylindrical (C), hook (H) and spherical (S) grasping activities. On the other hand, the remaining tasks lateral (L), tip (T) and palmar (P) grasping involve the use of precision and are grouped into a second group. As depicted in Fig. 2, in the proposed approach, the first stage pertains to the distinction on the basis of precision and power, which means that binary classification is carried out. Class indicators for all the power gestures is set equal to '1' and for the precision gestures are set to '0'.

The binary outputs are then collected and on training the data set again, the second stage is executed to classify the grasping activity as one of the three activities in that group. These predicted values are then used to compute the confusion matrix which gives the classification accuracy. Same is repeated for all the subjects in the complete data and the results are analyzed using statistical computations and displays. Categorization of the gestures now becomes a less complex task as compared to single stage method because the feature sets may be selected according to whether the distinction is to be made between power and precision or between the activities of one of these two groups. The PCA algorithm is used for reducing the dimensionality of the feature set so that the residual correlation does not interfere with the predicted values.

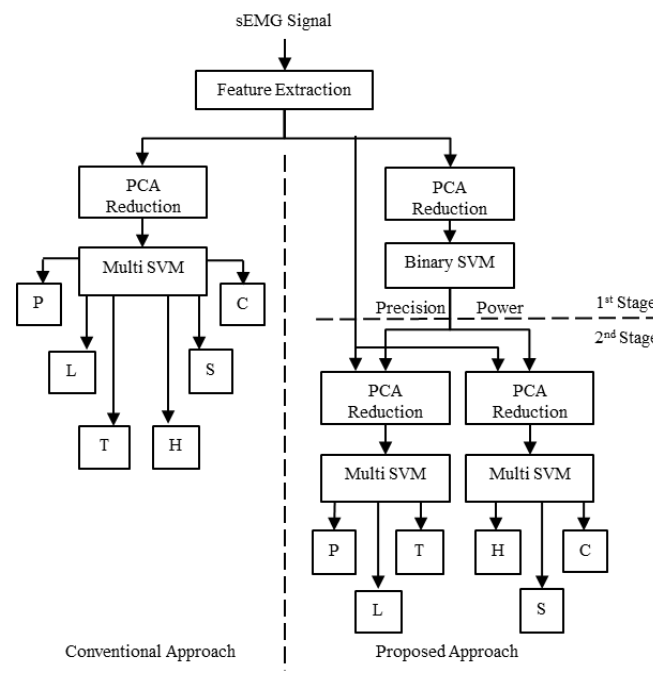

Fig. 2. Single stage classification vs. the proposed dual stage approach

IV. RESULTS AND DISCUSSION

In order to assess the categorization of the six activities in power and precision groups, a statistical analysis of their features is carried out using multivariate analysis of variance (MANOVA). This analysis is presented in Figure 3 in the form of a dendrogram tree. In a dendrogram, the U-shaped leaves represent the data point connections and the distance between two leaves gives the degree of dissimilarity, measured in dissimilarity index. Figure 3 displays the dendrogram tree for the 1st female subject plotted using the reduced feature set for the single stage classifier.

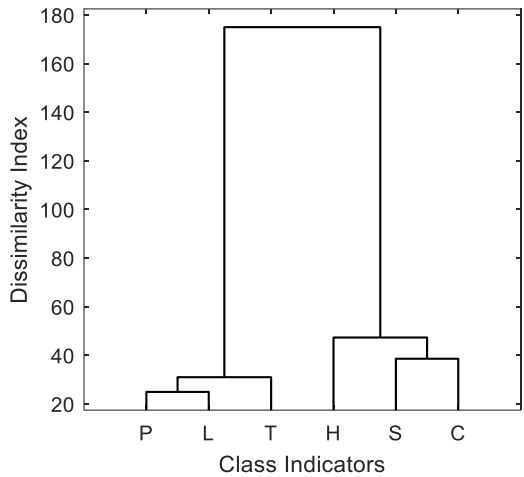

Fig. 3. Dendogram tree for the considered grasping activities

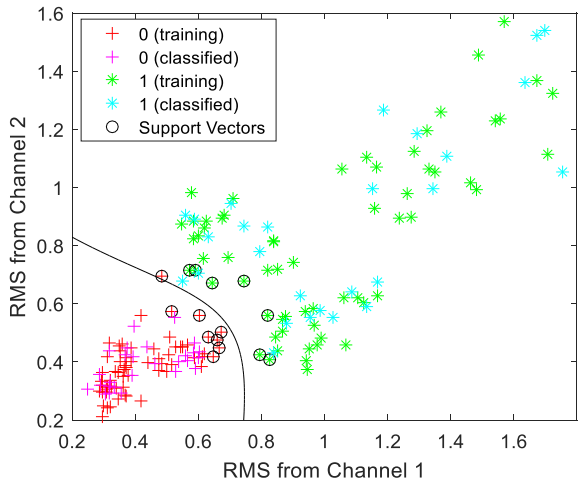

Fig. 4. Scatter plot of RMS from the two sEMG channels and the SVM classification in first stage of the proposed approach

From Fig. 3, it is evident that the three activities grouped under precision category namely, palmar (P), lateral (L) and tip (T) grasping are significantly distinct as compared to the other three activities hook (H), spherical (S) and cylindrical (C) grasping grouped under the power grasping category because the dissimilarity index is considerably higher between the two groups. Hence it is feasible to implement grasping activity classification using a dual-stage classification approach, since the categorizing an activity under the power or precision group is quite straight forward. To further demonstrate this argument, the RMS feature for all subjects determined from the two sEMG channels are plotted in Fig. 4.

The initial binary distinction phase groups solely on the basis of power (1) and precision (0). The degree of distinction for this purpose can be conveniently displayed using a scatter plot in Fig. 4 of the RMS feature. Evidently, a single feature is capable of accurately classifying a given activity under the power or precision group. Once the activity has been categorized as one of the two groups, the classification problem now reduces to identifying which of the three activities in the classified group has been performed.

The performance of the proposed dual-stage classification is compared with the conventional single stage classification in Table II and depicted in Fig. 5 for the individual subjects. Improved accuracy values for precision and power in Table II indicate a significant reduction in correlation. Accuracy values for the 1st stage validate the algorithmic process when compared to the single stage method accuracy values. Clearly, for each subject, the classification accuracy values have improved significantly for all the considered subjects when dual-stage classification is utilized. The combined accuracy for all the five subjects has also improved from 78.89% for single stage classification to 82.96% for dual-stage classification.

TABLE II. CLASSIFICATION ACCURACIES FOR CONVENTIONAL AND DUAL STAGE APPROACHES

| Stages | Proposed Dual Stage Method (%) | | | Single Stage Method (%) |
|---|---|---|---|---|
| Subjects | 1st Stage | 2nd Stage | Combined | |
| | | Precision | Power | | |
| Female-1 | 98.15 | 88.46 | 92.86 | 90.74 | 85.19 |
| Female-2 | 98.15 | 84.62 | 96.43 | 90.74 | 66.67 |
| Female-3 | 98.15 | 85.71 | 100 | 92.59 | 79.63 |
| Male-1 | 96.30 | 96.3 | 92.59 | 94.44 | 85.19 |
| Male-2 | 100 | 96.3 | 100 | 98.15 | 92.60 |
| Combined | 97.04 | 92.13 | 74.83 | 82.96 | 78.89 |

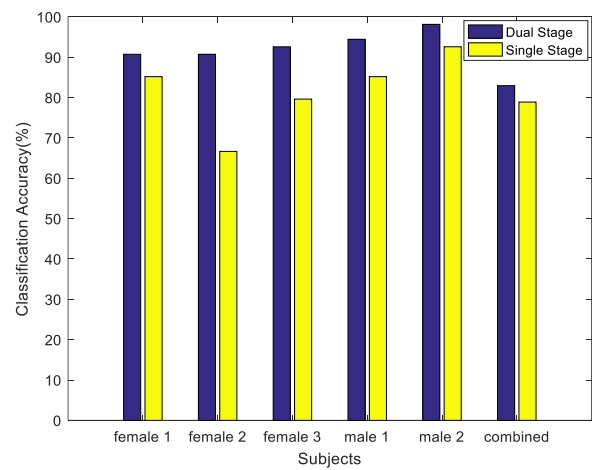

Fig. 5. Bar graph depicting the classification accuracies of the conventional single stage vs. the proposed dual-stage classification

V. CONCLUSION

In order to produce an efficient algorithm capable of assessing motion activities, a clear distinction in these gestures is of utmost importance. In this work, the hand grasping gestures are first classified as power or precision grasping. Statistical analysis is presented to validate the grouping of the activities used here. In the second stage of classification, the activity is then only to be classified from among the activities present in that group. The proposed dual stage classification provides a significant improvement in the classification accuracies for all subjects when compared to the conventional single stage classification. Feature reduction is performed using PCA for each stage of classification and the SVM classifier yields the predicted values. For the considered data, the accuracy of distinction increased from 78.89% for the conventional approach to 82.96% for the proposed dual stage approach. Thus, the algorithm presents itself as a suitable candidate for prosthetic development and hand gesture analysis.

ACKNOWLEDGMENT

The authors would like to recognize the funding support provided by the Science & Engineering Research Board, a statutory body of the Department of Science & Technology (DST), Government of India, SERB file number ECR/2016/000637.